\RequirePackage{fix-cm}

\documentclass[pdflatex]{sn-jnl}

\usepackage[numbers]{natbib}
\usepackage{cleveref}
\usepackage{physics}
\usepackage{mathrsfs}

\jyear{2022}%

\theoremstyle{thmstyleone}%
\theoremstyle{thmstyletwo}%

\theoremstyle{thmstylethree}%

\begin{document}

\title[Quantum Dynamics as a Ground State Problem with Neural Quantum States]{An Empirical Study of Quantum Dynamics as a Ground State Problem with Neural Quantum States}

\author*[1]{\fnm{Vladimir} \sur{Vargas-Calderón}}\email{vvargasc@unal.edu.co}

\author[1]{\fnm{Herbert} \sur{Vinck-Posada}}

\author[2]{\fnm{Fabio A.} \sur{González}}

\affil*[1]{\orgdiv{Grupo de Superconductividad y Nanotecnología, Departamento de Física}, \orgname{Universidad Nacional de Colombia}, \orgaddress{\postcode{111321}, \city{Bogotá},  \country{Colombia}}}

\affil[2]{\orgdiv{MindLab Research Group, Departamento de Ingeniería de Sistemas e Industrial}, \orgname{Universidad Nacional de Colombia}, \orgaddress{\postcode{111321}, \city{Bogotá},  \country{Colombia}}}


\abstract{We consider the Feynman-Kitaev formalism applied to a spin chain described by the transverse field Ising model.
This formalism consists of building a Hamiltonian whose ground state encodes the time evolution of the spin chain at discrete time steps.
To find this ground state, variational wave functions parameterised by artificial neural networks--also known as neural quantum states (NQSs)-- are used.
Our work focuses on assessing, in the context of the Feynman-Kitaev formalism, two properties of NQSs: expressivity (the possibility that variational parameters can be set to values such that the NQS is faithful to the true ground state of the system) and trainability (the process of reaching said values).
We find that the considered NQSs are capable of accurately approximating the true ground state of the system, i.e., they are expressive enough ansätze.
However, extensive hyperparameter tuning experiments show that, empirically, reaching the set of values for the variational parameters that correctly describe the ground state becomes ever more difficult as the number of time steps increase because the true ground state becomes more entangled, and the probability distribution starts to spread across the Hilbert space canonical basis.}




\maketitle

\section{Introduction}
A central problem of quantum physics, be it fundamental quantum physics or applications for quantum technology, is the ground state problem.
It can be defined as finding a state vector $\ket{\Psi}$ that minimises the expected value of the Hamiltonian $\hat{H}$ that represents the energetic interactions between the different parts that make up a quantum physical system.
It is well-known that the difficulty of solving the ground state problem for a physical system arises from the exponential growth of the Hilbert space with respect to the number of the system components and their dimension.
Therefore, techniques such as exact diagonalisation of $\hat{H}$ quickly render insufficient to find the ground state, and other approximate methods have to be used.

Interestingly, other central problems of quantum physics such as finding the evolution of a quantum system can be cast into the ground state problem, as demonstrated by the Feynman-Kitaev formalism~\citep{kitaev2002classical}.
An immediate implication of using this formalism is that the computational tools historically developed for solving the ground state problem can be used to find the dynamics of a physical system.
Broadly speaking, the Feynman-Kitaev formalism appends a clock as an auxilliary subsystem of the main physical system, i.e. the Hilbert space $\mathscr{H}$ of the whole system is $\mathscr{H} = \mathscr{P}\otimes\mathscr{C}$, where $\mathscr{P}$ is the Hilbert space of the main physical system and $\mathscr{C}$ is the Hilbert space of the clock.
It is possible to build a Hamiltonian $\hat{\mathcal{H}}:\mathscr{H}\to\mathscr{H}$ whose ground state $\ket{\Psi} \propto \sum_t \ket{\psi(t)}\otimes\ket{t}$ encodes the time evolution history of the main physical system, where $\ket{\psi(t)}$ its state at time $t$, and $\ket{t}$ is the state of the clock labelling time $t$~\citep{mcclean2013feynman}.
Therefore, getting the state of the physical system at a particular time $t$ can be done straight-forwardly by projecting the ground state $\ket{\Psi}$ onto the clock state $\ket{t}$, i.e. $\ket{\psi(t)} \propto \braket{\Psi}{t}$ (only the clock part is projected).

Recently, \citet{barison2022variational} showed how to compute the ground state of they Feynman-Kitaev associated Hamiltonian for a spin chain described by the transverse-field Ising model (TFIM) using a variational wave function based on variational quantum circuits, also known as parameterised quantum circuits. They mapped this Feynman-Kitaev Hamiltonian $\hat{\mathcal{H}}$ to a qubit Hamiltonian $\hat{\mathcal{H}}_Q$ with the same spectrum, and they found its ground state using the variational quantum eigensolver~\citep{kandala2017hardware} (VQE).
VQE consists of using a quantum computer to build a circuit composed of rotation gates whose angles are parameters that can be optimised.
Such a circuit might be written as $V(\vec{\vartheta})$, where $\vec{\vartheta}$ are the gate parameters.
Then, the circuit prepares the normalised quantum state $\ket{\phi_{\vec{\vartheta}}} = V(\vec{\vartheta})\ket{0,\ldots,0}$, where the state $\ket{0,\ldots,0}$ is the trivial all-zeros state that is normally used when initialising a quantum circuit.
After preparing the quantum state, it is used to measure the exact variational energy $E_{\vec{\vartheta}} = \expval{\hat{\mathcal{H}}_Q}{\phi_{\vec{\vartheta}}}$, and also to measure its derivatives with respect to each variational parameter.
Then, a classical computer is used to update the parameters, given an optimisation routine such as stochastic gradient descent.
However, we emphasise that the quantum circuit of VQE is simulated on a classical computer, which enables the exact access to the variational state $\ket{\phi_{\vec{\vartheta}}}$.
Therefore, the variational energy $E_{\vec{\vartheta}}$ is not estimated, as it would be on a real quantum device, but can be computed exactly.
The same occurs with the gradients of the variational energy with respect to variational parameters.
On a real quantum device, however, these quantities have to be estimated, which is both time-consuming and introduces inaccuracy.
Moreover, scalability of VQE to study large spin chains might be endangered by trainability issues in variational quantum circuits~\citep{anschuetz2022beyond} such as the onset of barren plateaus~\citep{mcclean2018barren}.

The aforementioned limitations of the standard VQE motivate us to approach the Feynman-Kitaev Hamiltonian through one of the most successful methods to solve the ground state problem: variational Monte Carlo (VMC)~\citep{becca2017quantum}, which aims to solve the problem $\min_{\vec{\theta}} \expval*{\hat{H}}{\Psi_{\vec{\theta}}}/\braket{\Psi_{\vec{\theta}}}$ (for any Hamiltonian $\hat{H}$), where $\ket{\Psi_{\vec{\theta}}}$ is a variational wave function, parameterised by some parameters $\vec{\theta}$.
VMC stands out because it does not compute the variational energy exactly; instead, VMC estimates the variational energy in a computationally efficient manner, by taking advantage of the fact that Hamiltonians that describe local interactions tend to follow an area-law scaling for the entanglement~\citep{eisert2010} (with notable exceptions~\citep{Vitagliano2010}), which ultimately means that only a small subset of elements of the Hilbert space basis is needed to accurately characterise the ground state (see more details about VMC in~\cref{sec:nqs}).

Of course, it is not obvious how to propose such a variational wave function $\ket{\Psi}_{\vec{\theta}}$. Recently, it has been shown that an outstanding parameterisation of variational wave functions can be achieved by bringing tools from the machine learning community; in particular, wave functions can be parameterised by artificial neural networks, giving birth to the so-called neural quantum states (NQSs)\footnote{Many of the different methods associated to NQSs have even been standardised in open-source libraries such as NetKet~\citep{netket3:2021}, which facilitates the use of these tools for researchers. All of the experiments of this paper involving VMC were done using NetKet.}~\citep{carleo2017solving}. 
Unfortunately, this means that the open questions from artificial neural networks also permeate their application to quantum physics.
In particular, there are two main areas of concern: expressivity and trainability.
Expressivity refers to the capacity that a parameterisation has to reproduce arbitrary wave functions~\citep{montufar2011expressive,gu2019representational,schuld2021effect}.
More precisely, a parameterised wave function defines a subset of all the possible wave functions; the larger this subset is, the more expressive the parameterisation is~\citep{sharir2021neural,sun2022entanglement}.
On the other hand, trainability refers to the capacity that an algorithm has to update the parameters so that a cost function--the expected value of $\hat{H}$--is minimised, taking into account the intricacy of said cost function~\citep{xiao2020disentangling}.
We emphasise that the present study aims to characterise expressivity and trainability of NQSs in the Feynman-Kitaev setting, but there are powerful alternatives such as direct integration of equations of motion for the variational parameters in NQSs through real time-evolution methods based on Monte Carlo sampling~\citep{freitas2018,Gutierrez2022realtimeevolution,schmitt2020,reh2021,donatella2022}.
Nevertheless, these are found to require exponentially many parameters to represent the quantum state at a given accuracy with respect to time~\citep{lin2022}, and suffer from numerical instabilities~\citep{hofmann2022}, especially near dynamical quantum phase transitions.

Limitations in expressivity and trainability in machine learning models used as NQSs also limit the possibility to successfully find ground states of Hamiltonians.
Therefore, it is imperative to understand what features found in a Hamiltonian can expose such limitations.
Therefore, the purpose of this work is to study the feasibility of an alternative to compute dynamics of quantum systems; that alternative being expressing the ground state of the associated Feynman-Kitaev Hamiltonian through an NQS.

We provide a systematic study of trainability and show that training NQSs through VMC to find the ground state of the Hamiltonian $\mathcal{H}$ is particularly difficult because the clock's degrees of freedom entangle with the main physical system, making the ground state $\ket{\Psi}$ not only highly entangled, but also in need of a large portion of the canonical basis of the Hilbert space $\mathscr{H}$ to be described.

This paper is divided as follows. In~\cref{sec:fk} we introduce the Feynman-Kitaev formalism. In~\cref{sec:nqs} we present the NQSs used in our work and explain how VMC works. Then, we present the results, which include an extensive study of VMC and NQSs hyperparameters, in~\cref{sec:results}. We discuss our empirical findings in~\cref{sec:discussion}. Finally, we conclude in~\cref{sec:conclusions}.

\section{The Feynman-Kitaev formalism}\label{sec:fk}

\begin{figure}[t]
    \centering
    \includegraphics[width=\textwidth]{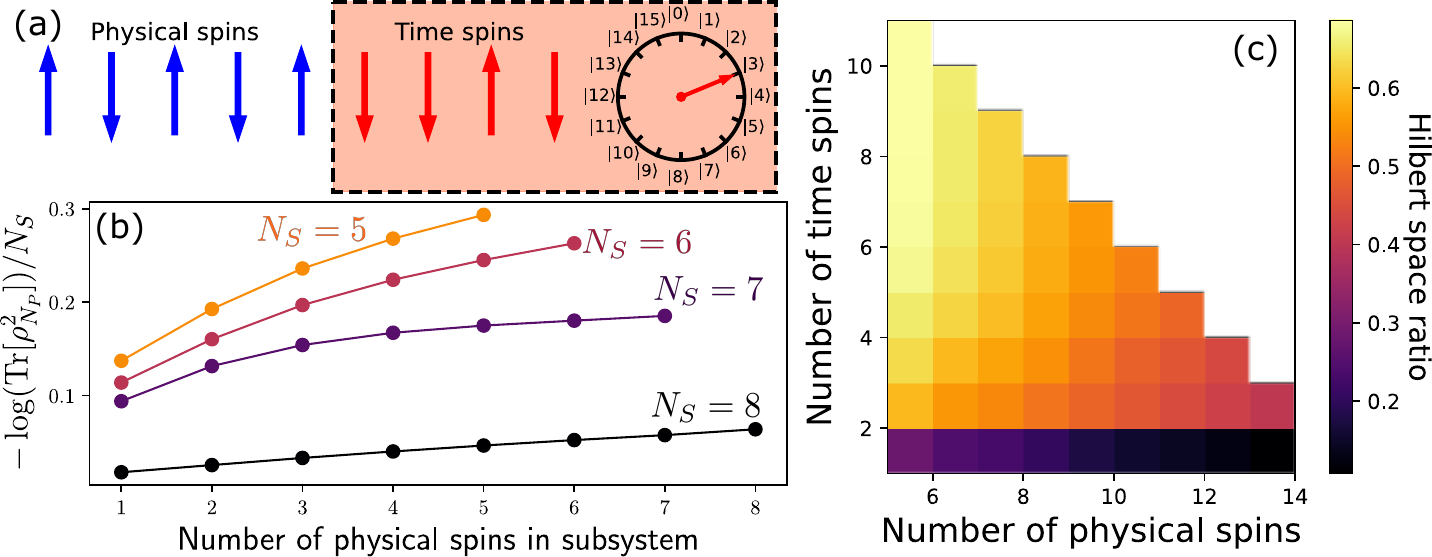}
    \caption{Representation of the physical spin chain enlargement with a clock state (a); and properties of the ground state of the enlarged Hamiltonian in~\cref{eq:bigHamiltonian} found with exact diagonalisation.
    (b) shows the second Rényi entropy per spin of a sub-chain of physical spins and (c) shows the ratio of the canonical basis of the Hilbert space that is needed to explain 99\% of the probability of the ground state. The main text (\cref{sec:results}) explains these plots in-depth.}
    \label{fig:gsProperties}
\end{figure}

One of the earliest proposals for performing a quantum computation was precisely that of the time evolution of a quantum system~\citep{Feynman85quantum}.
The idea behind this proposal is that the quantum state of a physical system can be described along with the quantum state of a clock~\citep{mcclean2013feynman}.
In particular, the clock can be in the states $\ket{0}, \ket{1},\ldots\ket{N}\in\mathscr{C}$, i.e., it is a $N+1$-level system in the Hilbert space $\mathscr{C}$.
In this paper, we encode the $N+1$-level system that describes the clock in $N_T = \lceil\log_2(N+1)\rceil$ spins\footnote{The purpose of this encoding is to have common ground with the study by~\citet{barison2022variational} for benchmarking reasons, but a true $N+1$-level system can be used instead.} using the reflected binary code (also known as Gray encoding) to map a state $\ket{t}$ to the state of $N_T$ spins because in this code, two consecutive states $\ket{t}$ and $\ket{t+1}$ are mapped to states of $N_T$ spins where only one spin is different.
Thus, the physical spin chain is enlarged with spins representing the state of the clock, as shown in~\cref{fig:gsProperties}(a).

\citet{mcclean2013feynman} showed that a variational principle can help in constructing a Hamiltonian $\hat{\mathcal{H}}$ of the physical + clock system such that its ground state is precisely
\begin{align}
    \ket{\Psi} = \frac{1}{\sqrt{N+1}}\sum_{t=0}^N \ket{\psi(t)}\otimes\ket{t}.\label{eq:historyState}
\end{align}

The Hamiltonian of the enlarged spin chain system can be written as~\citep{mcclean2013feynman,caha2018clocks,barison2022variational}
\begin{align}
    \hat{\mathcal{H}} = \hat{\mathcal{H}}_0 + \frac{1}{2}\sum_{t=0}^{N-1}\left[I_{\mathscr{P}}\otimes (\ketbra{t} + \ketbra{t+1}) + \{\hat{U}(\Delta t)\otimes\ketbra{t+1}{t} + \text{h.c.}\}\right],\label{eq:bigHamiltonian}
\end{align}
where $I_{\mathscr{P}}$ is the identity of the physical system's Hilbert space $\mathscr{P}$ and $\hat{\mathcal{H}}_0 = \hat{H}_0\otimes\ketbra{0}$ is a term that breaks the degeneracy of the ground state by fixing the initial state at $t=0$.
For instance,~\citet{barison2022variational} take $\hat{H}_0 = I_{\mathscr{P}} - \ketbra{\psi(0)}$, for any desired initial state of the physical system $\ket{\psi(0)}$.
Remarkably, the ground state of~\cref{eq:bigHamiltonian} is exactly~\cref{eq:historyState}, and its energy is $\expval*{\hat{\mathcal{H}}}{\Psi}=0$.
This property is important because it allows us to quantify how close the algorithm is to converge to the true ground state of the system.
It is worth highlighting that, even though whatever Hamiltonian $\hat{H}$ under consideration is sparse, $\hat{U}(T)$ need not be sparse, especially for large values of $T$~\citep{alhambra2022quantum}.
Therefore, computing the matrix representation of $\hat{U}(T)$ involves a considerable computational effort.
It is much easier to compute $\hat{U}(\Delta t)$ for $\Delta t = T/N \ll T$ for large $N$, as $\hat{U}(\Delta t)$ in~\cref{eq:bigHamiltonian} becomes just a perturbation of the identity.

In this work, we study the dynamics of the prototypical TFIM Hamiltonian defined on a one-dimensional chain of $N_S$ (ordered) spins:
\begin{align}
    \hat{H} = J\sum_{i=1}^{N_S - 1} \hat{\sigma}^z_{i}\hat{\sigma}^z_{i+1} + h\sum_{i=1}^{N_S}\hat{\sigma}^x_{i},\label{eq:tfim}
\end{align}
where $\hat{\sigma}^{z(x)}_{i}$ is the Z(X) Pauli operator acting only on spin $i$. Throughout the paper, we use $J=0.25$ and $h=1$.

\section{Variational Monte Carlo and neural quantum states}\label{sec:nqs}
The variational Monte Carlo (VMC) method leverages the variational method of quantum mechanics to problems with intractable Hilbert spaces~\citep{becca2017quantum}.
The variational method proposes a parameterised ansatz $\ket{\Psi_{\vec{\theta}}}$ and poses the problem $\min_{\vec{\theta}}\expval{\hat{\mathcal{H}}}{\Psi_{\vec{\theta}}}/\braket{\Psi_{\vec{\theta}}}$, where the energy is minimised, under the principle that it is only the ground state that has the minimum energy possible.
The problem is often difficult due to the non-convexity of the function to be minimised.
However, it is the dimensionality of the problem that renders it prohibitive to solve.
Indeed, by considering the completeness relation $I_{\mathscr{H}}=\sum_{\sigma}\ketbra{\sigma}$ of a Hilbert space $\mathscr{H}$ with an orthonormal basis $\{\ket{\sigma}\}$ ($\ket{\sigma}$ is a spin configuration--in the case of the TFIM expanded with a clock--of the explicit form $\ket{\sigma} = \ket{\sigma_1}\otimes\ket{\sigma_2}\otimes\cdots\otimes\ket{\sigma_{N_S+N_T}}\equiv\ket{\sigma_1,\ldots,\sigma_{N_S+N_T}}$ with $\ket{\sigma_i}\in\{\ket{\uparrow}, \ket{\downarrow}\}$), we have that the variational energy is
\begin{align}
    E_{\vec{\theta}} = \frac{\expval{I_{\mathscr{H}}\hat{\mathcal{H}}I_{\mathscr{H}}}{\Psi_{\vec{\theta}}}}{\expval{I_{\mathscr{H}}I_{\mathscr{H}}}{\Psi_{\vec{\theta}}}} = \sum_{\sigma,\sigma^\prime} P_{\vec{\theta}}(\sigma) \bra{\sigma}\hat{\mathcal{H}}\ket{\sigma^\prime}\frac{\Psi_{\vec{\theta}}(\sigma^\prime)}{\Psi_{\vec{\theta}}(\sigma)},\label{eq:ExactVariationalEnergy}
\end{align}
where $P_{\vec{\theta}}(\sigma) = \abs{\Psi_{\vec{\theta}}(\sigma)}^2/\sum_{\sigma^\prime}\abs{\Psi_{\vec{\theta}}(\sigma^\prime)}^2$ is the probability of the configuration $\sigma$.
The usual minimisation of $E_{\vec{\theta}}$ with respect to the parameters of the wave function can be performed using any optimisation algorithm.
However, as mentioned, it is practically impossible to perform the double summation because of the size of the Hilbert space $\mathscr{H}$.
Instead, VMC estimates $E_{\vec{\theta}}$ by virtue of the empirical fact that for many local-interaction Hamiltonians, $P_{\vec{\theta}}(\sigma) \approx 0$ for almost every $\sigma$, except for a small subset of the basis.
Therefore, the estimation of the variational energy is simply~\citep{becca2017quantum,carleo2017solving}
\begin{align}
    E^\star_{\vec{\theta}} = \expval{\sum_{\sigma^\prime} \bra{\sigma}\hat{\mathcal{H}}\ket{\sigma^\prime}\frac{\Psi_{\vec{\theta}}(\sigma^\prime)}{\Psi_{\vec{\theta}}(\sigma)}}_{\sigma\in\mathcal{M}},\label{eq:VariationalEnergy}
\end{align}
where the average is taken only using configurations from a sample $\mathcal{M}$ that is built according to the distribution $P_{\vec{\theta}}(\sigma)$.
Remarkably, since $\hat{\mathcal{H}}$ is usually sparse in the canonical basis $\{\ket{\sigma}\}$, the matrix elements $\bra{\sigma}\hat{\mathcal{H}}\ket{\sigma^\prime}$ are zero for most configurations $\sigma^\prime$ given a fixed $\sigma$. 
Another important feature of~\cref{eq:VariationalEnergy} is that the wave function need not be normalised to estimate the energy, or any other observable.
We also emphasise that, in the case of the Feynman-Kitaev Hamiltonian (\cref{eq:bigHamiltonian}) estimations of observables have to be multiplied by $N+1$ to account for the normalisation factor of the history state in~\cref{eq:historyState}.

\citet{carleo2017solving} introduced the idea of using neural networks to represent the wave function, i.e., the parameters $\vec{\theta}$ are the parameters of a neural network that takes as input a configuration $\sigma$ and outputs a complex number $\Psi_{\vec{\theta}}(\sigma)$.
These models receive the name of neural quantum states (NQSs).
A common choice of neural network is the restricted Boltzmann machine (RBM)~\citep{carleo2017solving}, which induces the ansatz:
\begin{align}
    \Psi_{\vec{\theta}}^{\text{RBM}}(\sigma) = \exp(\sum_{j=1}^{N_S+N_T} a_j\sigma_j) \prod_{\ell=1}^{N_H}2\cosh\left(b_\ell + \sum_{j=1}^{N_S+N_T} W_{\ell, j}\sigma_j\right),\label{eq:RBM}
\end{align}
where $N_H$ is the number of hidden units of the RBM and $\{a_j, b_j,W_{\ell,j}\}$ is the set of complex parameters.
The total number of parameters of this ansatz is $N_H(N_S+N_T + 1)$.

The sample $\mathcal{M}$ is built, in the case of the RBM, with the Metropolis-Hastings algorithm~\citep{hastings1970} because it is able to sample from a non-normalised distribution, such as the one induced by~\cref{eq:RBM}.
Indeed, normalising the RBM ansatz is computationally intractable for long spin chains.
The Metropolis-Hastings algorithm chosen in this study comprises the following steps: \textit{(i)} A random configuration $\sigma^{(0)}$ is generated.
\textit{(ii)} At iteration $r\ge 1$ we take the configuration $\sigma^{(r-1)}$ and randomly flip one spin, forming a candidate configuration $\tilde{\sigma}^{(r)}$.
\textit{(iii)} $\sigma^{(r)}$ is set to $\tilde{\sigma}^{(r)}$ with probability $\abs{\Psi^{\text{RBM}}_{\vec{\theta}}(\tilde{\sigma}^{(r)})}^2 / \abs{\Psi^{\text{RBM}}_{\vec{\theta}}(\sigma^{(r-1)})}^2$, else it is set to $\sigma^{(r-1)}$.
Usually, these steps are repeated until thermalisation, which means that the Markov chain stabilises, and only then one starts to collect configurations to build the sample $\mathcal{M}$.

However, stabilising Markov chains can be difficult, and in some cases might require a prohibitive amount of sampling in order to get a good representation of the probability distribution that needs to be approximated~\citep{vivas2022}.
For this reason, we also consider autoregressive models whose probability distribution can be sampled exactly, meaning that the sample $\mathcal{M}$ can be gathered by directly accessing the probability distribution $P_{\vec{\theta}}(\sigma)$.
In principle, avoiding the inherent practical problems of Markov chains for Monte Carlo sampling should bring an advantage; however, the performance of these autoregressive models did not meet these expectations.
We explain autoregressive models and report results based thereof in appendix~\ref{sec:autoregressive}.

\section{Results}\label{sec:results}

We find the ground state of~\cref{eq:bigHamiltonian}, which encodes the time evolution of a system governed by the TFIM Hamiltonian in~\cref{eq:tfim}.
The initial state is set as $\ket{\uparrow,\ldots,\uparrow}$, achieved by setting $\hat{H}_0=\frac{1}{2}\sum_{i=1}^{N_S}(1 - \hat{\sigma}^{z}_i)$.
There is an intrinsic difficulty in the Feynman-Kitaev Hamiltonian, which is that a lot of information (the quantum state of the Ising chain at each time step) needs to be stored in the ground state.
Such difficulty is evident from analysing the structure of the ground state of~\cref{eq:bigHamiltonian}, which we now denote by $\ket{\Phi(N_S,N_T)}$, where we explicitly denote the number of physical spins $N_S$ and the number of spins $N_T$ assigned to encode the temporal state.

Let us consider a system where the total number of spins is $N_T + N_S = 9$, which fixes the Hilbert space size to $\abs{\mathscr{H}} = 2^9$.
We can quantify the entanglement scale of the system by measuring the second Rényi entropy per physical spin $-\log(\Tr[\rho_{N_P}^2])/N_S$~\citep{torlai2018neuraltomography}.
Here, $\rho_{N_P}$ is the reduced density matrix of a sub-chain of $N_P$ physical spins, namely $N_P \le N_S$. 
This reduced density matrix is obtained by tracing over all the spin degrees of freedom except for the first $N_P$ spins. 
The second Rényi entropy is an entanglement quantifier and can be interpreted as follows: if a bipartite system is non-separable, when tracing the degrees of freedom of one part of the system, one is left with a reduced mixed state $\rho_{N_P}$ as a result; therefore, its decomposition will not have rank 1, and $\Tr[\rho^2_{N_P}]$ will be less than 1.
The more mixed the reduced density matrix, the smaller this trace will be, and the greater the second Rényi entropy will be.
\Cref{fig:gsProperties}(b) shows the second Rényi entropy by considering different physical spin sub-chains indicating that the entanglement increases between the first $N_P$ physical spins and the rest of the system as more spins are dedicated to encode time steps, and as we add more spins to the sub-chains.
Indeed, the last point of each curve in \cref{fig:gsProperties}(b) show that the entanglement between the time spins and the physical spins increase as long as more spins are used to encode the quantum state of the clock.

Another insightful analysis that summarises the complexity of the ground state of~\cref{eq:bigHamiltonian} is the proportion of the canonical basis elements needed to capture 99\% of the probability distribution given by the ground state $\ket{\Phi(N_S, N_T)}$ for different values of physical spins $N_S$ and temporal spins $N_T$, shown in~\cref{fig:gsProperties}(c).
The larger this ratio is, the larger the Monte Carlo sample $\mathcal{M}$ should be in~\cref{eq:VariationalEnergy} to be able to describe the expected energy to a given degree of error.
More formally stated, let the canonical basis $\{\sigma\}$ be indexed such that $\abs{\braket{\sigma^{(i)}}{\Phi(N_S,N_T)}}^2 \ge \abs{\braket{\sigma^{(i+1)}}{\Phi(N_S,N_T)}}^2$.
Then, \cref{fig:gsProperties}(c) shows the ratio $r/2^{N_S+N_T}$, where $r$ is the smallest integer such that $\sum_{i=1}^r \abs{\braket{\sigma^{(i)}}{\Phi(N_S,N_T)}}^2 > 0.99$.
It is clear from \cref{fig:gsProperties}(c) that for a fixed number of total spins $N_S+N_T$, the larger $N_T$ is, the highest the ratio of elements in the canonical basis needed to explain the ground state is.

\begin{figure}
    \centering
    \includegraphics[width=\textwidth]{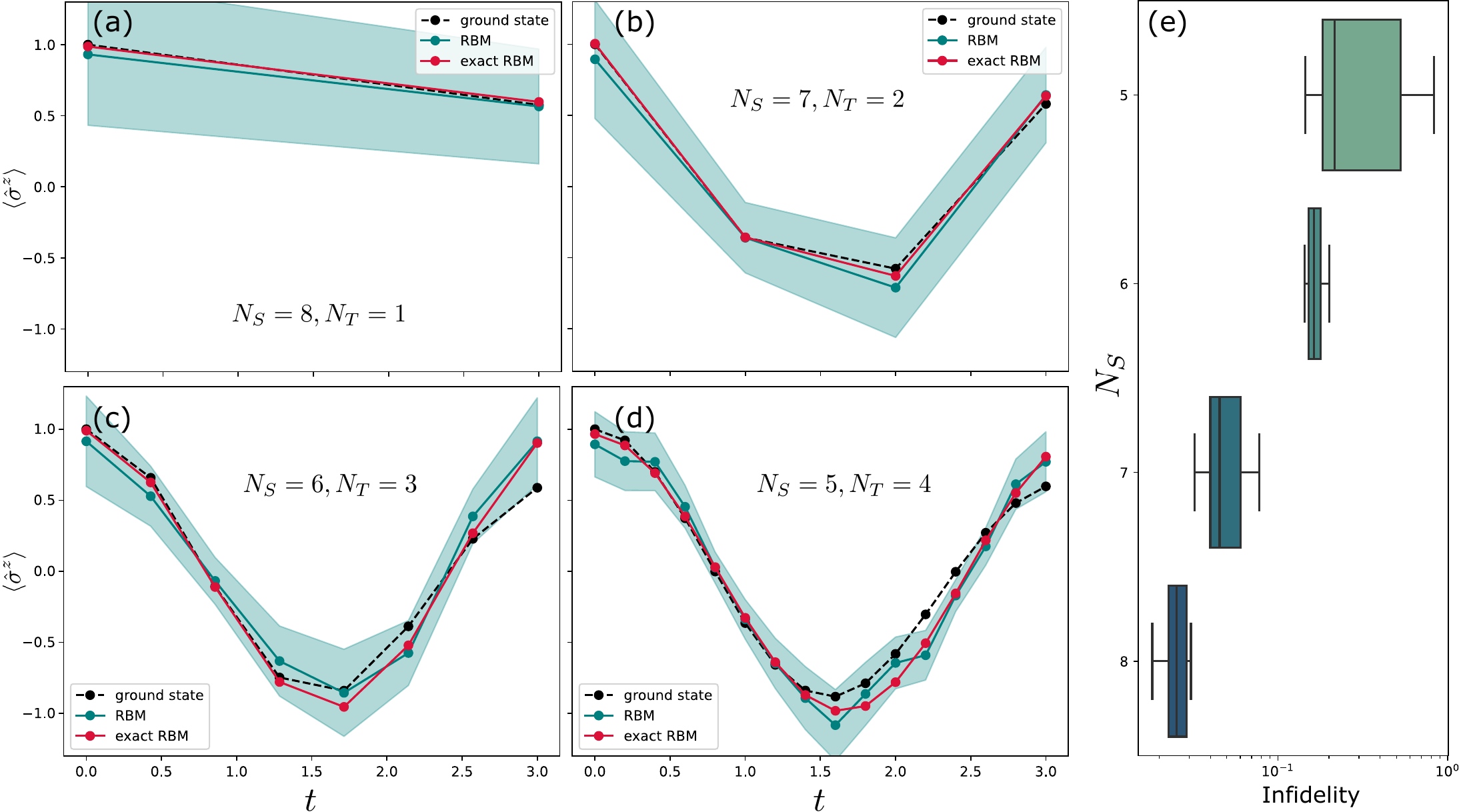}
    \caption{Time evolution approximated with an RBM as an NQS. (a)-(d) show the expected value of the average magnetisation $\expval{\hat{\sigma}^z}=\frac{1}{N_S}\sum_{i=1}^{N_S}\expval{\hat{\sigma}_i^z}$. In each panel, curves are shown for the average magnetisation obtained through exact diagonalisation (ground state), estimation of the variational magnetisation with a sample $\mathcal{M}$ (RBM) and exact variational magnetisation (exact RBM), which results from using the complete state vector instead of a sample. The shaded region indicates the estimated fluctuations of magnetisation using the sample $\mathcal{M}$. The lines serve as a guide for the eye only. (e) shows a box plot of infidelity ($1-\abs{\braket{\Phi(N_S,N_T)}{\Psi^{\text{RBM}}_{\vec{\theta^\star}}}}^2$) for the best 10 hyperparameter experiments, where $\vec{\theta}^\star$ indicates that parameters have been optimised until convergence.}
    \label{fig:RBMTimeEvolution}
\end{figure}

\Cref{fig:gsProperties} showed that the ground state, for large values of $N_T$, a well-spread and highly entangled ground state forms.
\Cref{fig:RBMTimeEvolution} reflects this fact on the increasing difficulty of training an RBM as an NQS for the ground state through VMC as $N_T$ grows.
Again, we fixed the total number of spins $N_S + N_T = 9$, as in~\cref{fig:gsProperties}(b).
For each value of $N_T$ (between 1 and 4), we performed hyperparameter tuning for 100 iterations with Optuna~\citep{optuna2019}, aiming to minimise the variational energy in~\cref{eq:VariationalEnergy}.
Details of the optimisation and hyperparameter tuning can be found in appendix~\ref{sec:optimisation}.
\Cref{fig:RBMTimeEvolution}(a)-(d) show the time evolution of the spin chain for the hyperparameters that produced the smallest infidelities.

It is seen in~\cref{fig:RBMTimeEvolution}(a)-(d) that, overall, the evolution of the average magnetisation is in accordance to the average magnetisation obtained through exact diagonalisation for all the values of $N_T$.
In the plots, the exact RBM line refers to magnetisation measured using the complete state vector from the RBM, instead of estimating the magnetisation through a sample.
However, the qualitative agreement of magnetisation curves does not exhibit the difficulty of training the RBM as $N_T$ increases. Indeed, the infidelities for~\cref{fig:RBMTimeEvolution}(a)-(d) are 0.018, 0.032, 0.144 and 0.145, respectively.
The increasing infidelity, as $N_T$ grows, indicates that training becomes more difficult, despite the Hilbert space always having the same size.
However, these are only the best states found after hyperparameter tuning.
\Cref{fig:RBMTimeEvolution}(e) shows a box plot of the infidelities of the best 10 hyperparameter tuning states, where a clear trend appears: the larger $N_T$ is, the more difficult it is to find the correct ground state.

\section{Discussion}\label{sec:discussion}

In this section, we discuss the results so far presented in light of the recent study by~\citet{barison2022variational}.
Compared to VQE, as we saw in the previous section, VMC struggles with finding an accurate approximation of the true ground state, presenting infidelities at least one order of magnitude higher than infidelities reported by~\citet{barison2022variational}. Unlike VMC, VQE directly handles a normalised quantum state in the whole Hilbert space, and its parameterisation consists of local transformations that preserve the norm.
A natural set of questions that arise are: what is it that makes  NQSs have larger infidelities than VQE? Is it expressivity? Is it trainability?~\citep{Wright:20,abbas2021power}
If the NQS can represent the ground state of~\cref{eq:bigHamiltonian} with low infidelity, it means that the NQS is expressive enough, but trainability hampers the possibility of describing the correct ground state, as shown in~\cref{fig:RBMTimeEvolution}.

\begin{figure}
    \centering
    \includegraphics[width=\textwidth]{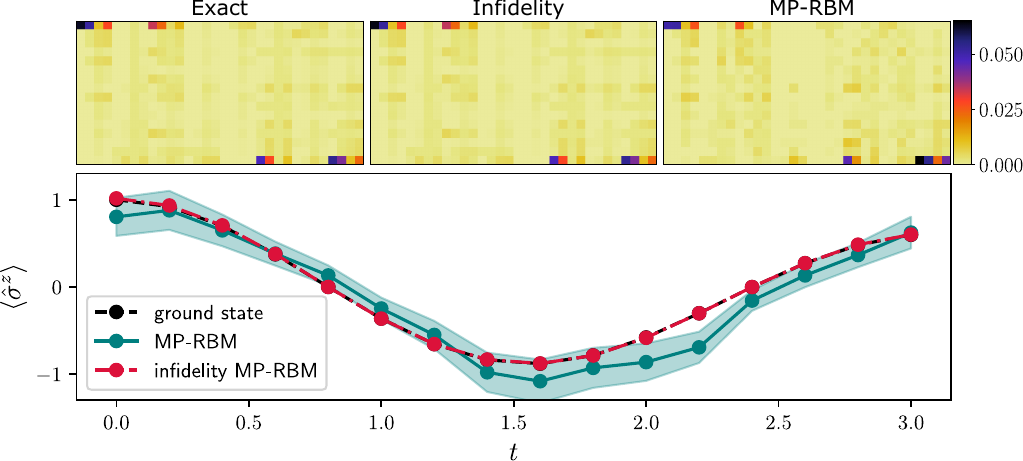}
    \caption{Probability of each element of the canonical basis of $\mathscr{H}$ and time evolution of magnetisation for an MP-RBM ansatz.
    The top panel shows the $2^9$ probabilities associated to each element of the canonical basis of $\mathscr{H}$ for the ground states obtained through exact diagonalisation, through variational minimisation of the infidelity, and through variational minimisation of the estimated energy in the left, middle and right sub-panels, respectively.
    The bottom panel shows the average magnetisation obtained with each of these states, where the ``ground state'' line corresponds to the magnetisation obtained with exact diagonalisation, the ``MP-RBM'' line is obtained through variational minimisation of the estimated energy, and the ``infidelity MP-RBM'' is obtained through variational minimisation of the infidelity.}
    \label{fig:MPRBMTimeEvolution}
\end{figure}

Considering the previous discussion, let us explore the expressivity of the RBM NQS.
The most challenging experiment tackled in this paper is the one of $N_S=5$ and $N_T=4$, which is perfectly tractable for a classical computer.
We consider the problem of finding parameters for the RBM ansatz that are able to faithfully describe the ground state of~\cref{eq:bigHamiltonian} by giving the RBM the ability to access the whole Hilbert space.
To this end, we directly minimise the infidelity $1 - \abs{\braket{\Phi(N_S=5,N_T=4)}{\Psi^{\text{RBM}}_{\vec{\theta}}}}^2$.
Experimentation with the ansatz in~\cref{eq:RBM} shows that training leads to local minima of the infidelity landscape, hinting convergence to stable excited states of~\cref{eq:bigHamiltonian}.
Therefore, we turned over to a similar RBM ansatz, which defines an RBM for the modulus and another for the phase of the wave function, namely the Modulus-Phase-RBM or MP-RBM~\citep{torlai2018neuraltomography}
\begin{align}
    \Psi_{\vec{\theta}}^\text{MP-RBM} = \exp(\Psi^{\text{RBM}}_{\vec{\theta}_{\Re}} + i \Psi^\text{RBM}_{\vec{\theta}_{\Im}}),\label{eq:RBMModPhase}
\end{align}
where $\Psi^{\text{RBM}}_{\vec{\theta}_{\Re}}$ and $\Psi^{\text{RBM}}_{\vec{\theta}_{\Im}}$ are RBMs defined by~\cref{eq:RBM}, with real-only parameters $\vec{\theta}_{\Re}$ and $\vec{\theta}_{\Im}$.
Training the MP-RBM ansatz in~\cref{eq:RBMModPhase} to minimise the estimated variational energy (see~\cref{eq:VariationalEnergy}) with VMC yields similar infidelities than the RBM ansatz after hyperparameter tuning (0.160 for the $N_S=5, N_T=4$ case).
However, it is easier to train the MP-RBM when minimising the infidelity (even without hyperparameter tuning).
In fact, we see that the MP-RBM is capable of learning the ground state with an infidelity of $2\times 10^{-3}$, as depicted by the excellent agreement between the MP-RBM magnetisation curve and the exact one in the bottom panel of~\cref{fig:MPRBMTimeEvolution}.

\Cref{fig:MPRBMTimeEvolution} exhibits the probability distribution of each state induced onto the canonical basis of the Hilbert space $\mathscr{H}$.
Since the infidelity-optimised MP-RBM leads to a very low infidelity, differences between its distribution (top middle panel) and the exact ground state distribution (top left panel) are minimal.
However, differences with the MP-RBM obtained through VMC are larger (cf.~\cref{fig:MPRBMTimeEvolution} top left and top right panels).
This is reflected onto the average magnetisation curves, shown in the bottom panel of~\cref{fig:MPRBMTimeEvolution}.
These results are in agreement with the study by~\citet{deng2017quantum}, who show that NQSs based on RBMs have a wide expressivity, able to represent many highly entangled quantum states.

This result supports the fact that NQSs are able to accurately approximate highly entangled ground states with widely-spread probability distributions.
The problem, however, resides on trainability when performing VMC.
An important final remark is that knowing the reason why infidelity optimisation consistently fails for NQSs with complex-only parameters remains an open question.

\section{Conclusions}\label{sec:conclusions}

We studied the time-evolution of a transverse field Ising model through the Feynman-Kitaev Hamiltonian, which encodes the state of a physical system at a given set of equidistant time instances into the state of an enlarged quantum mechanical system (it is enlarged by the state of a clock).
The ground state of the Feynman-Kitaev Hamiltonian was systematically searched by tuning hyperparameters of neural quantum state ansätze through the variational Monte Carlo method.

We showed that neural quantum states encounter difficulty in representing a highly-entangled ground state whose probability distribution is well-spread across the canonical basis of the Hilbert space.
As the number of clock states increased, we consistently saw that the performance of neural quantum states deteriorated, yielding lower fidelities to the true ground state of the Feynman-Kitaev Hamiltonian.
The characterisation of such ground state showed that as the number of clock states increased, both entanglement quantifiers and probability spread also increased.
These features explain that the ground state is ever more complicated for the neural quantum state to learn through variational Monte Carlo.

However, we also saw that the degrading performance of neural quantum states was not because of a lack of expressivity of the neural quantum state per se, as also supported by previous literature.
Instead, we provide evidence that trainability--in the variational Monte Carlo setup--is the main source of under-performance, even for autoregressive models, which sample directly from the probability distribution induced by the variational state.
This supports the hypothesis that optimisation techniques, and not sampling, degrade the quality of the learnt ground state, accompanied by the fact that energy convergence does not ensure convergence of the state (at least not in the same timescale)~\citep{vargas2020phase,ballentine2014quantum}.

\backmatter

\bmhead{Acknowledgments}

V. V.-C. and H. V.-P. acknowledge funding from Universidad Nacional de Colombia project HERMES 48528.

\section*{Declarations}

The authors declare no conflicts of interest. Further, data and code will be made available upon request to the authors.

\begin{appendices}

\section{Autoregressive ansatz}\label{sec:autoregressive}

We adopt a chain-type Bayesian network~\citep{brendan1998graphical,wu2019solving,wu2021unbiased,zhao2021overcoming} as the building block of the autoregressive model, which defines 
\begin{align}
    P_{\vec{\theta}}(\sigma) = q_{\vec{\theta}}(\sigma_1)\prod_{i=2}^{N_S+N_T} q_{\vec{\theta}}(\sigma_i \vert \sigma_1,\ldots,\sigma_{i-1}).\label{eq:autoregressiveProbModel}
\end{align}
Here, $q_{\vec{\theta}}(\sigma_i \vert \sigma_1,\ldots,\sigma_{i-1})$ models the probability that the $i$-th spin has a specified value conditioned on the observed values for the previous spins.
Thus, \cref{eq:autoregressiveProbModel} reads: the probability of the spin configuration $\sigma$ is the probability (assigned by the model parameterised with variational parameters $\vec{\theta}$) that the first spin has value $\sigma_1$, times the probability that the second spin has value $\sigma_2$ given that the first spin had value $\sigma_1$, and so on.
In other words, it is a straight-forward application of Bayes' rule.

Note that~\cref{eq:autoregressiveProbModel} is not concerned with the phase structure of the quantum state, i.e. $P_\theta(\sigma)$ is equivalent to $\abs{\Psi_\theta^\text{AR}(\sigma)}^2$.
In the same fashion, we can connect the wave function model to this probabilistic model through Born's rule, i.e. $q_{\vec{\theta}}(\sigma_i\vert\sigma_1,\ldots,\sigma_{i-1}) \equiv \abs{\Psi^{\text{AR}}_{\vec{\theta}}(\sigma_i\vert\sigma_1,\ldots,\sigma_{i-1})}^2$.
However, we need for an explicit model for $\Psi^{\text{AR}}_{\vec{\theta}}(\sigma_i\vert\sigma_1,\ldots,\sigma_{i-1})$ that complies with the aforementioned relations.
Thus, to describe the phase structure of the wave function, we express it as
\begin{align}
    \Psi^{\text{AR}}_{\vec{\theta}}(\sigma) = \prod_{i=1}^{N_S+N_T}\vec{\sigma}_i\cdot\vec{\eta}_{\vec{\theta}}(\sigma_i\vert \sigma_1,\ldots,\sigma_{i-1}),\label{eq:arnn}
\end{align}
where we have introduced the notation $\vec{\sigma}_i$ to denote the vector $(1, 0)^T$ for $\sigma_i = +1$ or $(0, 1)^T$ for $\sigma_i = -1$, and $\vec{\eta}_{\vec{\theta}}=(\eta_{\vec{\theta}}^{(1)}, \eta_{\vec{\theta}}^{(2)})^T$ denotes the needed complex vector generalisation of $q_{\vec{\theta}}$ such that $\abs{\eta_{\vec{\theta}}^{(1)}}^2+ \abs{\eta_{\vec{\theta}}^{(2)}}^2=1$, which ensures the correct normalisation of the probability model.

\begin{figure}
    \centering
    \includegraphics[width=\textwidth]{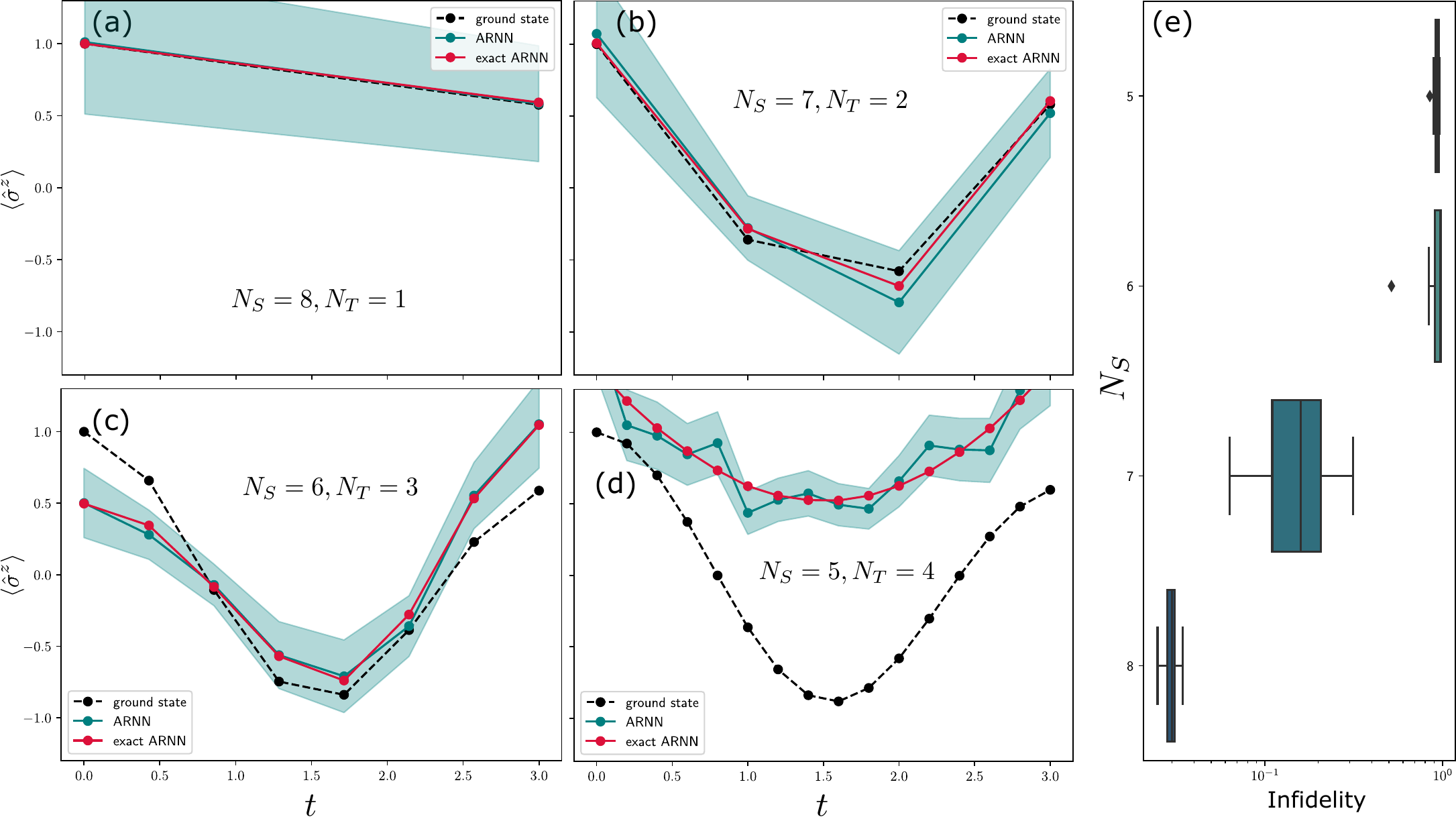}
    \caption{Similar to~\cref{fig:RBMTimeEvolution} but for the autoregressive ansatz in~\cref{eq:arnn}.}
    \label{fig:ARNNTimeEvolution}
\end{figure}

\Cref{fig:ARNNTimeEvolution} shows the evolution of magnetisation for the autoregressive ansatz in~\cref{eq:arnn}.
Qualitatively, results with this ansatz are similar to those of the RBM (cf.~\cref{fig:RBMTimeEvolution}), but they show worse performance in terms of correctly describing the evolution of magnetisation.
This is further confirmed by poor infidelities when $N_T$ is large.
The lowest infidelities achieved after hyperparameter tuning (see~\cref{sec:optimisation} for details) were 0.025, 0.063, 0.517 and 0.850 for~\cref{fig:ARNNTimeEvolution}(a)-(d), respectively.
Even though one can sample directly from the probability distribution induced by the autoregressive ansatz, avoiding issues with the Markov chain sampling, it is clear that capturing the ground state of~\cref{eq:bigHamiltonian} is more challenging for the autoregressive ansatz than the RBM ansatz.

From~\cref{fig:ARNNTimeEvolution} stands out the fact that, in some cases, the average magnetisation exceeds the upper bound limit for the average magnetisation, which is one.
This can be understood from the construction of the Feynman-Kitaev history state~\cref{eq:historyState}.
Explicitly, an observable $\hat{O}$ at time $t$ is measured as
\begin{align}
    O^*_{\vec{\theta}} = (N+1)\expval{\sum_{\sigma^\prime} \bra{\sigma}(\hat{O}\otimes \ketbra{t})\ket{\sigma^\prime} \frac{\Psi_{\vec{\theta}}(\sigma^\prime)}{\Psi_{\vec{\theta}}(\sigma)}}_{\sigma\in\mathcal{M}}.
\end{align}
Therefore, it is possible that the probability associated to a particular time of the clock is greater than $1/(N + 1)$, making it possible to measure average magnetisations greater than one.

Finally, we found that optimising the infidelity (instead of minimising the variational energy) for the ansatz in~\cref{eq:arnn} traps the NQS into an excited state of~\cref{eq:bigHamiltonian}, which is why we turned over to an autoregressive ansatz that explicitly divides the modulus and phase of the wave function, similar to the works by~\citet{hibat2020recurrent} and~\citet{barrett2022autoregressive}.
In this setup, we divide the autoregressive neural network $\vec{\eta}$ into two autoregressive neural networks, one for the modulus, and the other for the phase of the wave function.
Training this ansatz to minimise the estimated variational energy with VMC, yields high infidelity of 0.920 after hyperparameter tuning for the $N_S=5,N_T=4$ case, which is a similar infidelity to the one obtained by the ansatz in~\cref{eq:arnn}.
On the other hand, the infidelity optimisation (without any hyperparameter tuning) yields an infidelity of $3.5\times 10^{-3}$, comparable to that of the MP-RBM.
These findings are further evidence for NQSs being able to express highly entangled ground states with widely-spread probability distributions, pin-pointing trainability as the main problem for learning ground states.

\section{Optimisation of neural quantum states}\label{sec:optimisation}
Both VMC and NQS training have hyperparameters that dictate the behaviour of the variational energy optimisation.
Since the aim of this study is to train NQSs in the VMC setup with the greatest possible quality, we adopt a fruitful machine learning strategy that targets the best set of hyperparameters, namely hyperparameter tuning.
Hyperparameter tuning is a difficult meta-optimisation task that, in our case, thrives to answer the questions: what is the best structure of the training algorithm, and what is the NQS architecture that produces the lowest variational energy?

Let us start by stating the hyperparameters for VMC.
The two main components of VMC are the sampler and the optimiser.
The sampler dictates how the sample $\mathcal{M}$ of~\cref{eq:VariationalEnergy} is built, and the optimiser is a rule for updating the parameters $\vec{\theta}$ of the NQS.

In the NQS literature, it is common to find that stochastic reconfiguration (SR)~\citep{sorella2007weak} is used in combination with stochastic gradient descent (SGD) as an optimiser.
SR takes into account the geometry of the variational energy landscape to update parameters in the directions that yield maximum descent.
However, we experimented on optimisation instances that used the RBM NQS (\cref{eq:RBM}) with different numbers of hidden neurons using both SR+SGD and AdamW~\citep{adamw} and found no significant difference in performance.
On the contrary, AdamW was faster, which is why we chose it as the optimisation method for all of the experiments shown in the main text.
We consider its learning rate as the sole hyperparameter of the optimiser.
Regarding the sampler, we consider the number of parallel Markov chains and the number of total samples as its two hyperparameters.
In the case of an autoregressive NQS, no Markov chains are considered, and the sampler only has the number of total samples hyperparameter.

The hyperparameters for the architecture of the NQSs are different for the RBM and the autoregressive ansätze.
For the RBM, the hyperparameter is $\alpha:=N_H/(N_S+N_T)$, which specifies the proportion of hidden neurons with respect to the visible neurons of the RBM.
For the autoregressive ansatz, the autoregressive neural network $\vec{\eta}$ in~\cref{eq:arnn} has two hyperparameters: the number of layers $N_L$, and the number of hidden neurons $N_H$ of each layer, with the property that the layers are masked in such a way that the conditional probability of a spin taking a value depends only on the values of the previous spins.

\begin{table}
\caption{Prior distribution of hyperparameters for the tree-structured Parzen estimator.}\label{tab:hyperparams}
\begin{center}
\begin{tabular}{llll}
\toprule
NQS                  & hyperparameter         & Support & Prior \\
\midrule
\multirow{2}{*}{RBM \& AR}            & Number of samples       & $[256,2048]$        & Uniform      \\
  & Learning rate & $[10^{-4}, 1]$ & Loguniform\\
  \midrule
\multirow{2}{*}{RBM} & Number of Markov chains &    $\{4,8,16\}$     &    Uniform   \\
                     & $\alpha$                &    $\{1,\ldots,5\}$     & Uniform      \\
                     \midrule
\multirow{2}{*}{AR}  & $N_L$                   &   $\{1,2,3\}$      &    Uniform   \\
                     & $N_H$                   &     $\{2,4,8,16,32\}$    &     Uniform \\
\bottomrule
\end{tabular}
\end{center}
\end{table}
The hyperparameter tuning algorithm that we used is the tree-structured Parzen estimator (TPE)~\citep{bergstra2011algorithms} provided in the Optuna package~\citep{optuna2019}.
In summary, TPE works by jointly modelling the distribution $\ell(x)$ of features that have corresponding figures of merit below a given threshold $y^*$ and, similarly, the distribution $g(x)$ of features with corresponding figures of merit above said threshold.
The models $\ell$ and $g$ are tree-structured models with single-variable priors for each hyperparameter, which are shown in~\cref{tab:hyperparams}.
Hyperparameter tuning was conducted for 100 different hyperparameter sets for each ansatz, and for each combination of number of physical spins $N_S$ and number of time spins $N_T$.

\subsection{Turning on the clock adiabatically}
We adopt the strategy by~\citet{barison2022variational} of turning on the clock gradually.
This means that we perform the energy (or infidelity) minimisation of~\cref{eq:bigHamiltonian} for a total time $T_k = kT/20$, starting from $k=1$ and ending at $k=20$.
This strategy simplifies learning overall, as it gradually increases the learning problem difficulty: for small $k$, the evolution is for small times, meaning that the state of the physical system remains almost unchanged throughout evolution.
As $k$ gets larger, the state of the physical system starts to significantly change between consecutive time steps.




\end{appendices}


\bibliographystyle{plainnat}
\bibliography{apssamp}


\end{document}